# Understanding Polaronic Transport in Anatase TiO$_2$ Films by Combining Precise Synthesis and First-Principles Many-Body Theory

Fengdeng Liu[1,2,*], Zhifei Yang[1,3,*], Yao Luo[4,*], Silu Guo[1], Chi Zhang[5], Sooho Choo[1], Xiaotian Xu[5], Xiaojia Wang[5], Andre Mkhoyan[1], Marco Bernardi[4,a], and Bharat Jalan[1,a]

[1]Department of Chemical Engineering and Materials Science, University of Minnesota – Twin Cities, Minneapolis, Minnesota 55455, USA

[2]Department of Electrical and Computer Engineering, University of Minnesota – Twin Cities, Minneapolis, Minnesota 55455, USA

[3]School of Physics and Astronomy, University of Minnesota – Twin Cities, Minneapolis, Minnesota 55455, USA

[4]Department of Applied Physics and Materials Science, and Department of Physics, California Institute of Technology, Pasadena, California 91125, USA

[5]Department of Mechanical Engineering, University of Minnesota, Minneapolis, Minnesota 55455, USA

* Equally contributed

[a] Corresponding authors: bjalan@umn.edu ; bmarco@caltech.edu

**Abstract**

In complex oxides, charge carriers often couple strongly with lattice vibrations to form polarons–entangled electron-phonon quasiparticles whose transport properties remain difficult to characterize. Experimental access to intrinsic polaronic transport requires ultraclean samples, while theoretical descriptions demand methods beyond low-order perturbation theory. Here, we combine the growth of high-quality oxygen-vacancy–doped anatase $TiO_2$ films by hybrid molecular beam epitaxy (MBE) with a first-principles electron–phonon diagrammatic Monte Carlo (FEP-DMC) framework recently developed for accurate polaron predictions. Our films exhibit record-high electron mobility for anatase $TiO_2$, in excellent agreement with FEP-DMC calculations conducted prior to experiment, which predict a room-temperature mobility of $45 \pm 15$ $cm^2V^{-1}s^{-1}$ and a mobility-temperature scaling of $\mu \propto T^{-1.9 \pm 0.077}$. Microscopic analysis using scanning transmission electron microscopy and X-ray photoelectron spectroscopy reveals the role of oxygen vacancies in modulating transport at lower temperatures. FEP-DMC further provides quantitative insight into polaron formation energy, phonon cloud distribution, lattice distortion around the polaron, and the polaronic contribution to mobility. Together, these results establish a predictive theory-experiment workflow to characterize polarons and provide a microscopic understanding of large-polaron transport in anatase $TiO_2$, with broader implications for complex oxides and other polaronic materials.

**Introduction**

Polarons resulting from strong electron-phonon (e-ph) interactions are commonly found in inorganic materials with polar or ionic bonds. In a simple picture, the electron or hole charge carrier polarizes and distorts the surrounding lattice (*1*). This effect can be viewed as the formation of a phonon cloud accompanying the charge carrier. In the limit of strong e-ph interactions, the carrier becomes self-trapped at a lattice site, and transport occurs by thermally activated diffusion (*1*). The strength and length scale of the e-ph interactions provide criteria to classify polarons. Large polarons are spatially delocalized and originate from long-range e-ph interactions, while small (self-localized) polarons form in the presence of strong, short-range e-ph interactions (*1*). Polarons govern the spectral and transport properties in complex oxides (*2*), and thus a deeper understanding of polaron effects is important in materials physics. In particular, charge transport in polaronic materials can occur through a range of mechanisms, which have been studied extensively using Hamiltonian models, and more recently first-principles calculations (*1*), but are difficult to predict in real materials.

Among polaronic oxides, titanium dioxide ($TiO_2$) has attracted attention due to its wide band gap (*3, 4*), chemical stability (*3, 5*), optical absorption (*6-8*), and (photo)catalytic properties (*9-13*). $TiO_2$ has three different polymorphs: rutile, anatase, and brookite (*6, 14*). Anatase $TiO_2$ is a transparent conducting oxide that is known to exhibit pronounced polaron effects, as evidenced by the presence of broadened quasiparticle peaks accompanied by phonon satellites in angle-resolved photoemission spectroscopy (ARPES) spectra (*15*). Here, we use anatase $TiO_2$ as a prototypical oxide to show a combined theory-experiment approach for precise characterization of polaronic transport in the intrinsic, phonon-limited regime. This development is made possible by a combination of advances, respectively, in calculations of strong e-ph interactions and growth of high-quality oxide thin films.

A unified formalism to study e-ph interactions ranging from weak to strong is essential to model transport in polaronic materials (*1*). First-principles calculations using the conventional Boltzmann transport equation (BTE) approach are limited to materials with weak e-ph interactions, where electron spectral functions exhibit sharp quasiparticle peaks and polarons are absent (*16, 17*). Some of us have recently developed the first-principles e-ph diagrammatic Monte-Carlo (FEP-DMC) method, which quantitatively sums e-ph Feynman diagrams to all orders using material-specific e-ph interactions computed from first principles (*18*). FEP-DMC can make accurate predictions of charge transport for both small and large polarons in a unified framework (*18*). In parallel, hybrid molecular beam epitaxy (MBE) has emerged as a key approach for atomically-precise synthesis of oxides with low defect density (*19*).

Here, we study high-quality anatase $TiO_{2-\delta}$ thin films grown by hybrid MBE and carry out a systematic analysis of their structural and electrical properties, focusing on polaronic transport. Our mobility measurements demonstrate the scaling behavior between mobility $\mu$ and temperature as $\mu \propto T^{-1.993 \pm 0.048}$, in remarkable agreement with our FEP-DMC predictions – made prior to experiment – which point to the presence of large polarons with Frohlich character and a dominant coupling with optical phonons. These results underscore both the accuracy of the theoretical framework and the exceptional quality of the films. Oxygen vacancies act as intrinsic *n*-type dopants, supplying the carriers that enable transport in this system. By uniting predictive FEP-DMC calculations with hybrid MBE growth and quantitative transport measurements, our work establishes a general framework for microscopic understanding of polaronic charge transport. This combined theory-experiment strategy opens new avenues for unraveling and engineering polaronic phenomena across complex oxides and other quantum materials.

**Results and Discussion**

Anatase TiO$_2$ (tetragonal, $a = b = 3.78$ Å, $c = 9.51$ Å) can be grown epitaxially on LaAlO$_3$ (001) due to the pseudocubic structure of LaAlO$_3$ (001) (pseudocubic lattice parameter = 3.79 Å) and the small in-plane lattice mismatch of -0.2% (*20, 21*). Details of the growth process are provided in the Methods section. A metal-organic precursor, titanium tetraisopropoxide (TTIP), is used to supply both Ti and O and no additional oxygen source is used (*22, 23*). This is done deliberately to create oxygen vacancies in films to provide *n*-type carriers and induce conductivity. In **Fig. 1A**, the high-resolution X-ray diffraction (HRXRD) 2$\theta$-$\omega$ coupled scans of films grown at four different substrate temperatures ($T_{sub}$ = 500 °C, 600 °C, 700 °C, 900 °C) show that the TiO$_{2-\delta}$ films exhibit phase-pure anatase structures. Wide-angle scans of these films are shown in **Fig. S1A**. The inset in **Fig. 1A** shows the film thickness extracted from X-ray reflectivity measurements as a function of $T_{sub}$, where the film thickness increases and stabilizes with increasing substrate temperature. The trend of such $T_{sub}$-dependent growth rate is similar to that of rutile TiO$_2$ films grown by hybrid MBE (*22*). **Figure 1B** shows the *in-situ* reflection high-energy electron diffraction (RHEED) patterns along the [100]$_{sub}$ azimuthal direction after growth. The RHEED patterns are spotty when $T_{sub}$ = 500 °C and 600 °C, and become streaky when $T_{sub}$ is increased to 700 °C and 900 °C, suggesting different growth modes in these two temperature regimes (*22*). The streaky RHEED patterns at $T_{sub} \geq 700$ °C also indicate smooth surface morphology consistent with the surface step terraces in atomic force microscopy (AFM) and low root-mean square surface roughness as shown in the inset of **Fig. 1E**. At $T_{sub} \geq 700$ °C, RHEED images show a clear ¼ order diffraction peaks (marked by arrows in **Fig. 1B**), which agrees with the previous studies (*20, 24-* 26). We further show the measured room-temperature Hall mobility $\mu_{300K}$ as a function of growth temperatures in **Fig. S1B** along with the AFM images on these samples.

We also perform thermal conductivity measurements to characterize the structural quality of our films using time-domain thermoreflectance (TDTR). **Figure 1C** shows TDTR data for a representative 345 nm anatase $TiO_2$ film grown at 700 °C for 5 hours. To eliminate growth-induced oxygen vacancy defects, we anneal the film at 700 °C for 2 hours in an oxygen-rich atmosphere using a furnace. The annealed sample exhibits a two-terminal resistance exceeding 200 MΩ, confirming negligible oxygen vacancy concentration. The through-plane thermal conductivity ($\Lambda_z$) of the 345 nm anatase $TiO_2$ film is $4.3 \pm 0.4$ W m$^{-1}$ K$^{-1}$, comparable to reported values for bulk anatase single crystals along the [001] direction (*27, 28*). The in-plane thermal conductivity is determined using the beam-offset TDTR method (**Fig. S2**), yielding $\Lambda_x \approx \Lambda_y = \Lambda_r = 13.9 \pm 5.4$ W m$^{-1}$ K$^{-1}$. These results are consistent with prior calculations (*27*) and highlight a pronounced anisotropy between in-plane and out-of-plane thermal transport, reflecting the anisotropic crystal structure of anatase $TiO_2$.

To characterize the electronic transport, we introduce oxygen vacancies as intrinsic *n*-type dopants by performing an in situ post-growth vacuum annealing at $T_{sub}$ = 900 °C for 15 minutes. The thermal profile during growth and annealing is illustrated in **Fig. 1D**. Structural characterization by RHEED (inset, **Fig. 1D**), HRXRD (**Fig. 1E**) confirms that the resulting anatase $TiO_{2-\delta}$ films are phase-pure, epitaxial, and atomically smooth. A leftward shift of the film peak in XRD further indicates an increased oxygen-vacancy concentration accompanied by an expanded out-of-plane lattice parameter. Temperature-dependent Hall measurements using the van der Pauw geometry are then carried out to probe transport properties. The results, shown in **Figs. S3A–C**, include the carrier density $n_{3D}$, Hall mobility $\mu$, and antisymmetrized Hall resistance $R_{xy}$ as a function of magnetic field *B*. The negative slopes in $R_{xy}$ confirm *n*-type conduction due to oxygen vacancies, while no measurable conduction is observed in the substrates.

Across all films, the electron mobility follows a scaling relation $\mu \propto T^\alpha$. A representative $\mu$ vs. *T* curve (**Fig. 2**) shows very close agreement with FEP-DMC theoretical predictions conducted prior to the transport measurements. For context, mobility data from other studies (*29-32*) are also plotted. Across the full temperature range, hybrid MBE-grown films exhibit record-high electron mobilities – reaching 31 cm$^2$V$^{-1}$s$^{-1}$ at 300 K and 65 cm$^2$V$^{-1}$s$^{-1}$ at 200 K – approaching the phonon-limited lower bound predicted by FEP-DMC (27.5 cm$^2$V$^{-1}$s$^{-1}$ at 300 K and 70.2 cm$^2$V$^{-1}$s$^{-1}$ at 200 K). This demonstrates that, even in the presence of oxygen vacancies, charge transport is primarily limited by phonon scattering at $T \geq 200$ K. Data below 200 K are excluded due to contributions from the Kondo effect which is discussed in the following. Power-law analysis reveals a temperature trend $\mu \propto T^\alpha$, with $\alpha = -1.993 \pm 0.048$, in excellent agreement with the FEP-DMC prediction ($\alpha = -1.900 \pm 0.077$). Importantly, FEP-DMC calculations identify longitudinal optical (LO) phonons as the dominant scattering channel via the Fröhlich interaction (**Figs. S4A-B**). Furthermore, unlike the conventional Boltzmann transport equation (BTE), which systematically overestimates the mobility by neglecting polaronic effects, FEP-DMC incorporates higher-order e-ph interactions governing polaronic behavior and reproduces both the magnitude and temperature dependence of the experimental mobilities (**Fig. S4C**). A more detailed discussion of the FEP-DMC calculations is given in Section IV in Supplementary Information.

To investigate the microscopic structure of polarons and to resolve their energy dispersion in anatase TiO$_2$, **Fig. 3A** shows the computed FEP-DMC polaron dispersion band structure (orange line). The G$_0$W$_0$ electronic band structure is also shown for comparison (blue line). Energies are referenced to the conduction-band minimum at Γ. The polaron band structure is much flatter than the electronic band structure but is still dispersive, which shows the presence of large polarons in anatase TiO$_2$, whereas a dispersion-less band would indicate small polarons (*1*). We compute a mass enhancement factor of $m_\mathrm{p}/m_\mathrm{e} = 1.6$ along the Γ-X directions, where $m_\mathrm{p}$ is the polaron effective mass and $m_\mathrm{e}$ is the electron effective mass, consistent with the value of 1.7 from ARPES measurements (*15*). From Fig. 3A, we compute the polaron formation energy as the difference between the lowest polaron energy and the lowest electronic energy. Extrapolation to infinite system size gives a polaron formation energy of -0.178 eV, as reported in our previous work (*18*).

In **Fig. 3B,** we resolve the contribution of each phonon mode to the polaron state, where the marker (solid circle) size is proportional to the amplitude of each phonon mode, $Z_Q$, in the phonon cloud. This result shows that long-wavelength optical phonons give a dominant contribution, allowing us to classify the polaron in TiO$_2$ as a Fröhlich-like large polaron formed by the long-range Fröhlich interactions (*33*). The side panel, showing the momentum-integrated phonon spectral function, confirms the dominant role of high-energy optical modes. Because the FEP-DMC method returns the complete ground-state polaron wavefunction, we can characterize the phonon number distribution $Z_N$ (**Fig. 3C**) in the polaron cloud. The rapid decay beyond $N = 3$ shows that on average only a few phonons dress the charge carrier. The zero-phonon amplitude $Z_{N=0}$, which corresponds to the electron quasiparticle weight, is smaller than 1 but remains substantial ($Z_{N=0} \approx 0.6$), underscoring the intermediate-coupling nature of the large polaron. Note that our FEP-DMC transport calculations neglect the phonon-assisted current term; however, a Fermi's-golden-rule estimate shows that its contribution is negligible (see Section IV in Supplementary Information).

The atomic displacements induced by the polaron are typically not evaluated in DMC because this method preserves the translational symmetry (*34*). Here, by treating the polaron as a superposition of localized electron and phonon clouds centered at each unit cell, we develop a technique, to be detailed elsewhere, for computing the atomic displacements associated with the polaron in anatase $TiO_2$. For a many-body polaron wavefunction $|\Phi_0\rangle$, we introduce the e-ph distortion function

$$d_{n,\kappa\alpha}(R_p) = \left\langle \Phi_0 \left| \sum_{R_i} \hat{c}_n^\dagger(R_e)\hat{c}_n(R_e)\hat{u}_{\kappa\alpha}(R_p + R_e) \right| \Phi_0 \right\rangle$$

which quantifies the displacement of atom $\kappa$ along direction $\alpha$ at position $R_p$ relative to the center of the electronic Wannier orbital $n$. We develop an efficient diagrammatic sampling technique to calculate this e-ph distortion function. The resulting atomic displacements are shown in **Fig. 3D** and **3E**, in the *b-c* and *a-b* planes respectively. Because of the large-polaron character, the displacements extend over thousands of unit cells. Interestingly, the lattice perturbation induced by the polaron is more short-ranged along the *c* axis compared to the *a/b* directions, as seen by the more rapid decay of the displacements in the *c* direction. We attribute this anisotropy to the less dispersive band structure along the *c* direction ($\Gamma - Z$ direction in **Fig. 3A**) compared to the *a-b* plane band dispersion ($\Gamma - X$ direction in **Fig. 3A**). The anisotropic character is particularly important for surface polarons in $TiO_2$, where the surface termination is expected to impart different polaron energetics and dynamics.

We now turn to the low-temperature electrical transport properties of anatase $TiO_{2-\delta}$ films, focusing on two key features: (1) a pronounced in-plane resistance anisotropy, and (2) an upturn in resistance below ~200 K. Using 4-terminal van der Pauw measurements (**Fig. S5A-E**) and XPS spectroscopy (**Fig. S5F** and **Fig. S6**), we can attribute these features to the Kondo effect induced by the $Ti^{3+}$ states arising from oxygen vacancies (See Section V in Supplementary Information for detailed analysis). Scanning transmission electron microscopy (STEM) provides direct evidence linking oxygen vacancies to the observed transport anisotropy. The low-magnification HAADF-STEM image of a 69 nm $TiO_{2-\delta}$/$LaAlO_3$ (001) film grown at 700 °C (**Fig. 4A**)–the sample exhibiting large in-plane resistance anisotropy (**Fig. S5C**)–shows nonuniform contrast across the film. An atomic-resolution image of the white-box region (right panel, **Fig. 4A**) confirms the anatase structure, consistent with HRXRD results. Higher-resolution imaging (**Fig. 4B**) further reveals spatial variations: some regions display a perfect anatase lattice (rectangle 1), while others exhibit alternating dark-bright striping along the $[\bar{2}03]$ direction (rectangle 2), indicative of oxygen-vacancy ordering (*35*). Line-scan analysis of HAADF images (**Fig. 4C**) further confirms these periodic intensity variations (~2.1 nm spacing), providing direct evidence of a heterogeneous vacancy distribution. At higher vacancy densities, diffusion across the film homogenizes this distribution, explaining the reduced anisotropy observed at elevated growth temperatures. Collectively, these findings show that both the in-plane anisotropy and the low-temperature resistance upturn in anatase $TiO_{2-\delta}$ films arise from oxygen vacancies – through their density, spatial inhomogeneity, and associated $Ti^{3+}$ states. Future spectroscopy studies will aim to resolve the microscopic origin of the observed contrast.

**Conclusions**

In summary, we demonstrate a unified theory–experiment framework for understanding charge transport in the polaronic regime by combining FEP-DMC calculations and hybrid MBE synthesis of high-quality anatase $TiO_{2-\delta}$ thin films. Our measurements reveal record-high mobilities and scaling behavior in quantitative agreement with predictions made prior to experiment, highlighting the robustness of FEP-DMC for describing large-polaron transport. Detailed FEP-DMC analysis provides microscopic insight into the polaron state beyond the reach of experiments, including the number and type of phonons in the polaron cloud and the atomic displacements and lattice distortion associated with the polaron. Beyond polaronic conduction, we establish that oxygen vacancies act as intrinsic *n*-type dopants and critically influence low-temperature transport phenomena, including in-plane resistance anisotropy and the Kondo effect. These conclusions are supported by complementary XPS, STEM, and transport measurements, which directly link vacancy density and spatial distribution to anisotropic conduction and magnetic scattering. Together, these results advance the understanding of anatase $TiO_2$ as a model polaronic system, improve its electronic performance, and introduce a broadly applicable workflow for probing and engineering polaronic transport with atomic resolution in complex oxides and quantum materials.

**Methods**

**Anatase $TiO_2$ film growth.** All $TiO_2$ thin films were grown on 5 mm × 5 mm × 0.5 mm $LaAlO_3$ (001) substrates using a hybrid metal-organic Molecular Beam Epitaxy approach (*22, 23*). Titanium and oxygen were supplied via a metal−organic precursor, titanium tetraisopropoxide (TTIP, 99.999% pure; Sigma-Aldrich). TTIP vapor entered the chamber through a heated gas injector (E-Science, Inc.) in an effusion cell port that was in direct line-of-sight with the substrate. The TTIP vapor reached the injector via a linear leak valve followed by a capacitance manometer (Baratron, MKS Instruments, Inc.), the valve opening serving as the control variable and the manometer pressure serving as the process variable in a control loop. The substrate temperatures were varied from 500 °C to 900 °C. An *in-situ* post-growth annealing was performed at 900 °C for 15 minutes for some films as outlined in **Fig. 1D** of the main text.

**Film Characterization.** A Rigaku SmartLab XE diffractometer was used for X-ray diffraction measurements. Film thicknesses were extracted from the X-ray reflectivity (XRR). X-ray photoelectron spectroscopy (XPS) was used to determine the Ti fraction and valence. To determine compositions of the films, XPS fine scans were measured using a Physical Electronics 5000 VersaProbe III photoelectron spectrometer with monochromatic Al Kα X-rays, a 55-eV analyzer pass energy, a 50-ms time step and a normal emission geometry. The spectra of C 1$s$ and Ti 2$p_{3/2}$ were fitted with Lorentzian-Gaussian functions after subtracting the background signal and the C-C peak was calibrated to 284.8 eV. Then the $Ti^{3+}$ contents of the films were determined by calculating the peak area ratios of $Ti^{3+}$ and $Ti^{4+}$, defined as $A(Ti^{3+})/A(Ti^{3+} + Ti^{4+})$.

**Scanning transmission electron microscopy (STEM).** The cross-sectional samples for STEM study were prepared using Focused Ion Beam (FIB) (FEI Helios NanoLab G4 dual-beam). A 50 nm of amorphous carbon was first deposited onto the film by thermal evaporation. An additional 2 μm thick amorphous carbon layer was deposited on the region of interest using Ga ion beam to protect this region during FIB milling. The FIB ion beam was operated at 30 kV. STEM imaging was conducted on an aberration-corrected FEI Titan G2 60–300 (S)TEM microscope equipped with a CEOS DCOR probe corrector, monochromator, and a super-X energy dispersive X-ray (EDX) spectrometer. The STEM was operated at a voltage of 200 kV. All the HAADF-STEM images were acquired using a 90 pA probe current. The camera length was set to 130 mm with a probe convergence angle of 25.5 mrad. The HAADF detector's collection angles ranged from 55 mrad to 200 mrad.

**Thermal transport measurements.** The anisotropic thermal conductivities of a representative anatase TiO$_2$ film were measured with the ultrafast pump-probe technique, time-domain thermoreflectance (TDTR), along both the through-plane (routine TDTR) and in-plane (beam-offset) directions (*36-38*). The TiO$_2$ film was grown at 700 °C for 5 hours, followed by an *ex-situ* furnace annealing at 700 °C for 2 hours in an oxygen-rich atmosphere. The film thickness is estimated to be 345 nm based on the calibrated growth rate. Before thermal measurements, a 65-nm Al film was deposited on the TiO$_2$ film, the LaAlO$_3$ substrate, and a SiO$_2$ (300-nm)/Si reference sample, serving as the transducer. The thickness of the Al transducer, determined from picosecond acoustics, is consistent for all samples. For routine TDTR measurements of the TiO$_2$ through-plane thermal conductivity ($\Lambda_z$), a 5× objective lens (with a 1/$e^2$ radius of ~12 *m*m) and two modulation frequencies of 9 and 18 MHz were used. The through-plane thermal conductivity of the TiO$_2$ film was extracted by comparing the experimental data to a 3D heat transfer model (*36*). For the in-plane thermal measurements with the beam-offset approach, the TiO$_2$ film was studied using a 50× objective lens (1/$e^2$ radius of ~1.3 μm) and a modulation frequency of 1.5 MHz to enhance the measurement sensitivity (*37, 39, 40*). The thermal conductivities in two orthogonal in-plane directions (namely, $\Lambda_x$ and $\Lambda_y$) were measured twice for sample loaded at 0° and 90°. The in-plane thermal conductivity is determined by fitting the full width at half maximum (FWHM) of the out-of-phase signal ($V_{out}$) to the model simulation (*39, 41*).

**Electrical transport measurements**. Electrical transport measurements were performed in a Quantum Design Dynacool physical property measurement system (PPMS) in a van der Pauw geometry with aluminum wire bonding to connect samples to the resistivity measurement puck. The temperature range is between 1.8 K and 400 K, and the magnetic field is between ±9 T.

**DFT calculations.** DFT calculations in anatase $TiO_2$ employ the Quantum Espresso package with the PBESol exchange-correlation functional (*42, 43*) and norm-conserving pseudopotentials (*44, 45*). We use a plane-wave kinetic energy cutoff of 85 Ry and a lattice parameter $a$ = 3.78 Å with the ratio $c/a$ = 2.51. We use the Yambo (*46*) code to compute the $G_0W_0$ corrected band structure for the lowest 10 conduction bands. The $G_0W_0$ calculation includes 500 bands to compute the polarization function with the Bruneval-Gonze terminator (*47*). We verified that our $G_0W_0$ band structure in anatase $TiO_2$ is consistent with previous work (*48*). The Wannier functions are constructed using a 6×6×6 using *k*-point grid with the Wannier90 code (*49*). We conduct density functional perturbation theory calculations on a *q*-point grid of 6×6×6 to compute the lattice dynamics and e-ph perturbation potential, and calculate the e-ph interactions using Perturbo (*50*).

**Acknowledgements**

Synthesis (F.L.) is supported primarily by the National Science Foundation (NSF) through the Future of Semiconductor (FuSe) grant under award number DMR-2328702, and under award number DMR- 2306273. Transport measurements were supported by the Air Force Office of Scientific Research (AFOSR) through grants FA9550-21-1-0025, and FA9550-23-1-0247. Film growth was performed using instrumentation funded by AFOSR DURIP awards FA9550-18-1-0294 and FA9550-23-1-0085. Part of the work (X-rays spectroscopy) was supported by the U.S. Department of Energy (Award No. DE-SC0020211). Z.Y. acknowledges partial support from the UMN MRSEC program under Award No. DMR-2011401. S.G. and K.A.M. were supported partially by the UMN MRSEC program under Award No. DMR-2011401. This work also benefitted from the AFOSR Multi University Research Initiative (MURI, Award No. FA9550-25-1-0262). Parts of this work were carried out at the Characterization Facility, University of Minnesota, which receives partial support from the NSF through UMN MRSEC. Portions of this work were carried out at the Minnesota Nano Center, which receives support from the NSF through the National Nanotechnology Coordinated Infrastructure (NNCI) under Award No. ECCS-2025124. Y.L. and M.B. were supported by the National Science Foundation under Grant No. OAC-2209262. Y. L. acknowledges support from the Eddleman Fellowship. Thermal studies (C.Z., X.X., and X.W.) were supported by the UMN MRSEC program under Award No. DMR-2011401.

**Author Contributions:** F.L., Z.Y., M.B. and B.J. conceived the idea and designed the experiments. F.L. grew the films. Z.Y. developed the fabrication process. F.L. and Z.Y., performed structural characterization and electrical testing. STEM/EDX measurements were performed by S.G. under the supervision of K.A.M. The first-principles calculations were performed by Y.L. under the supervision of M.B. TDTR measurements were done by C.Z. and X.X. under the supervision of X.W. F.L., Z.Y., Y.L, M.B. and B.J. wrote the manuscript. All authors contributed to the discussion and manuscript preparation.

**Competing Interest Statement:** The authors declare no competing interests.

**Data and materials availability:** All data needed to evaluate the conclusions of the paper are present in the paper and/or the Supplementary Materials.

**Supporting Information**

The supplementary information contains additional details on the structural characterization using XRD and AFM of anatase $TiO_{2-\delta}$ films. Details of electrical transport measurements, XPS measurements and the FEP-DMC calculations are included.

**Figures and Tables**

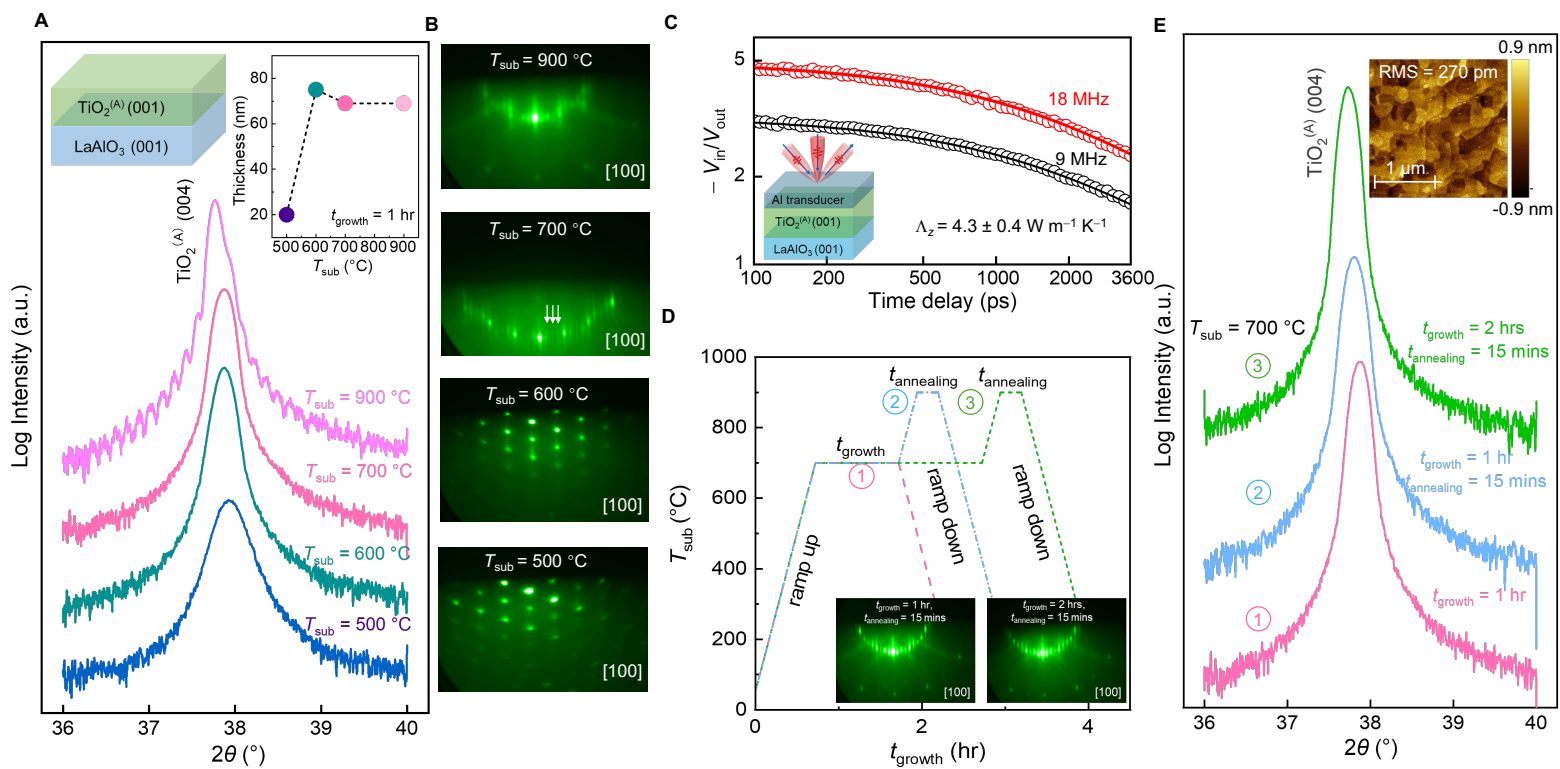

**Figure 1. Characterizations of anatase TiO$_{2-\delta}$ films.** **(A)** High resolution X-ray diffraction $2\theta$-$\omega$ coupled scans of TiO$_{2-\delta}^{(A)}$ (001)/LaAlO$_3$ (001) thin films with varying substrate temperatures $T_{sub}$. The insets show the schematic of the film structure (left) and the thickness of the films as a function of $T_{sub}$. **(B)** Reflection high-energy electron diffraction (RHEED) patterns of these thin films with the beam pointing along the [100] direction. **(C)** Time-domain thermoreflectance ratio (TDTR) signals at room temperature with simultaneous two-frequency fittings on a representative 345 nm film. The inset shows the schematic of the sample structure and the measurement process. **(D)** Outlines of the growth process and *in-situ* annealing process. The insets show RHEED patterns for films with the annealing process where the beam points along the [100] direction. **(E)** High resolution X-ray diffraction $2\theta$-$\omega$ coupled scans of TiO$_{2-\delta}^{(A)}$ (001)/LaAlO$_3$ (001) thin films outlined in (D). The inset shows the atomic force microscopy (AFM) image of the sample 1 outlined in (D) and root mean square (RMS) of the surface roughness.

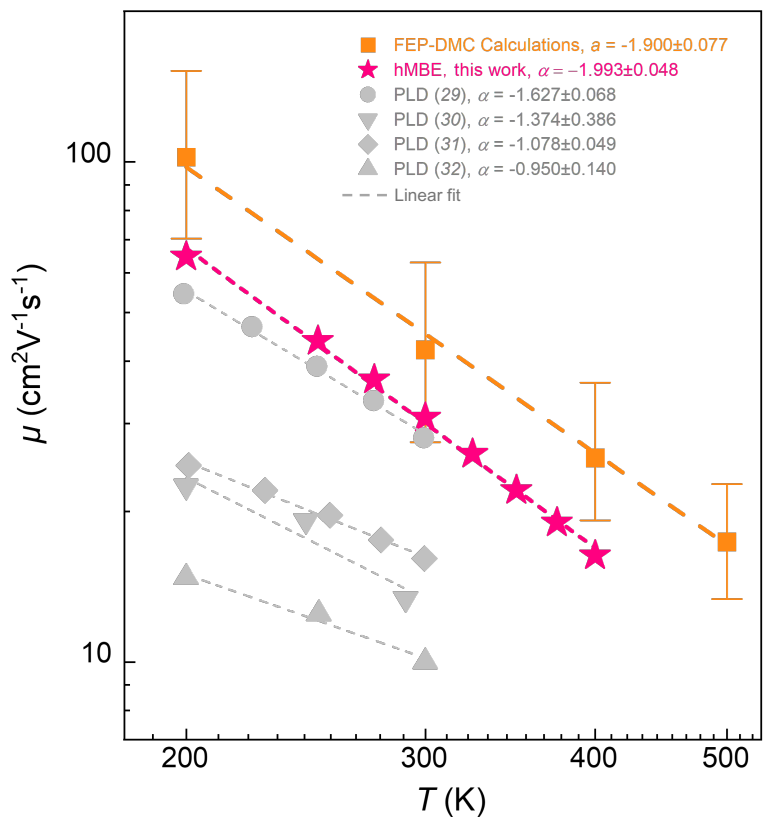

**Figure 2. Comparison of carrier mobility in anatase TiO$_2$.** Comparison of the computed phonon-limited carrier mobility using the FEP-DMC method (orange squares) with the measured temperature-dependent electron mobility in anatase TiO$_{2-\delta}$ film grown by hybrid MBE (pink stars). Experimental mobility data from previous studies are shown with gray symbols for comparison (*29-32*).

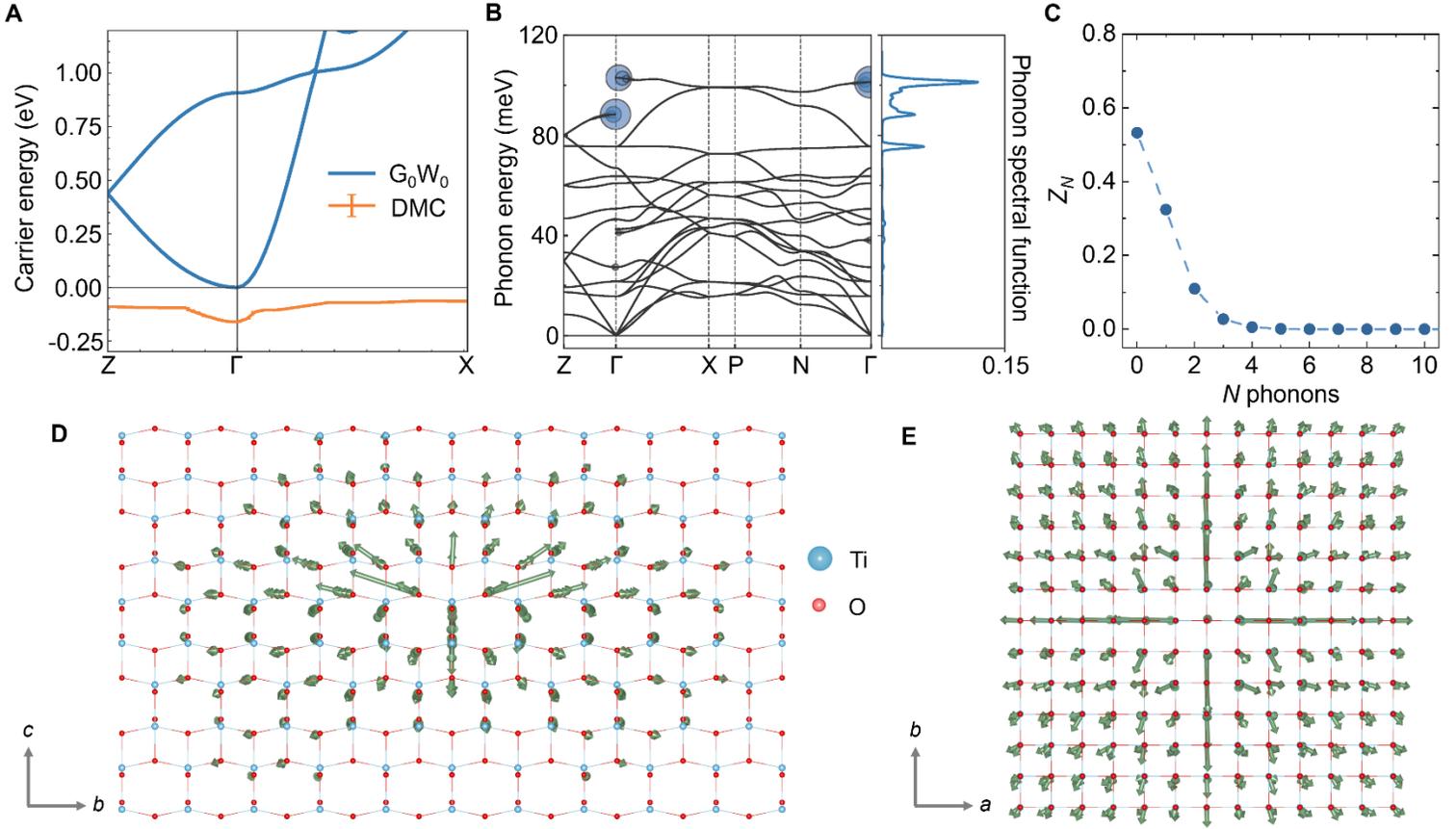

**Figure 3. Theoretical modeling of large polarons in anatase TiO$_2$.** **(A)** Comparison of the computed FEP-DMC polaron dispersion and the G$_0$W$_0$ electronic band structure. Energies are referenced to the conduction-band minimum at Γ. The DMC error bars are smaller than the linewidth. **(B)** The contribution of each phonon mode, $Z_Q$, in the polaron ground state along high-symmetry lines in the Brillouin zone. The side panel shows the phonon spectral function, $\sum_Q Z_Q \delta(\omega - \omega_Q)$. **(C)** The phonon number distribution $Z_N$ obtained from the polaron wavefunction. Atomic displacements associated with the polaron, shown for the O atoms in the **(D)** b-c and **(E)** a-b planes respectively. The displacements are obtained from the e-ph distortion function defined in the text using FEP-DMC. The length of the arrows is scaled up by 1000 times for visualization purposes.

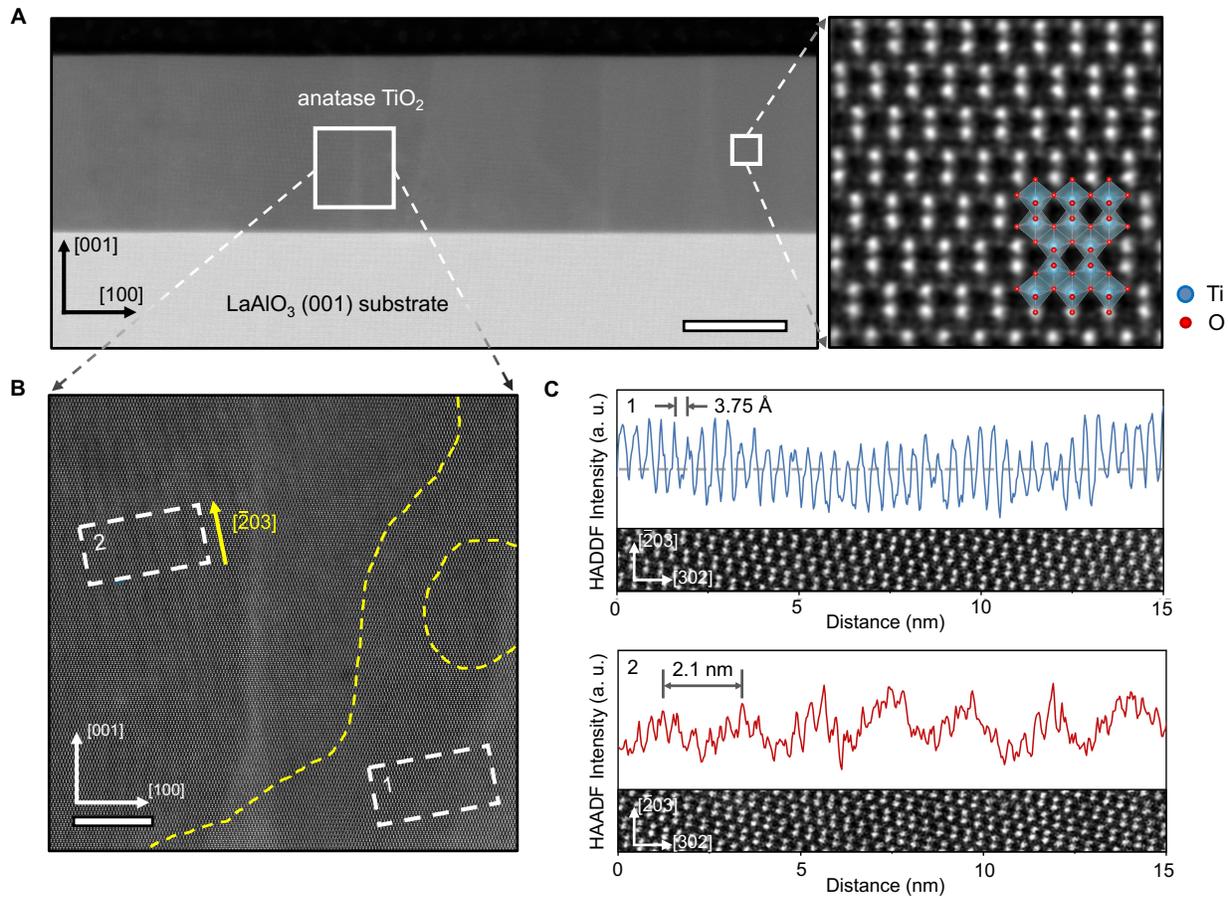

**Figure 4. Oxygen vacancies in anatase TiO$_{2-\delta}$ film**. **(A)** Low-magnification HAADF-STEM image of an anatase TiO$_{2-\delta}$ film grown at 700 °C on an LaAlO$_3$ (001) substrate with a thickness of ~70 nm. The atomic-resolution STEM image of the anatase structure (right) is taken at the white box region. Scale bar = 50 nm. **(B)** A selected region from the anatase TiO$_2$ thin film, showing dark-bright strip features along the $[\bar{2}03]$ crystallographic direction. Yellow dashed line marks the boundary between the region with strips and the region with uniform intensity. Scale bar = 10 nm. **(C)** HAADF intensity line scans from regions 1 and 2 in (B). The line scan from region 1 shows a constant average intensity along the grey dashed line, suggesting a uniform region without oxygen vacancies. The inset atomic-resolution image shows the $[\bar{2}03]$ out-of-plane direction crystal structure. The line scan from region 2 exhibits periodic peaks in addition to the smaller oscillations, where the valleys correspond to dark-strip areas with oxygen vacancies (oxygen deficiencies), while the peaks represent the normal TiO$_2$ intensity from the bright-strip areas. The spacing between the peaks is approximately 2.1 nm, indicating the distribution of oxygen vacancies. All the images are low-pass filtered for better visualization.